\documentclass[aps,prd,prabib,showkeys,nofootinbib,10pt]{revtex4}
\usepackage{graphicx} \usepackage{amsmath} \usepackage{amssymb}
\usepackage{amsfonts} \usepackage{bm}
\usepackage{array}
\usepackage{siunitx}
\usepackage[singlelinecheck=false]{caption}
\usepackage{ytableau}
\usepackage{multirow}
\usepackage{mathtools}
\usepackage{fancyhdr}

\usepackage{tikz}

\usepackage{xcolor}

\fancypagestyle{specialfooter}{%
\fancyhf{}

\fancyfoot[R]{ \noindent\fbox{%
\parbox{\textwidth}{%
{\footnotesize \bf \tiny{This version of the article has been accepted for publication, after peer review (when applicable) but is not the Version of Record and does not reflect post-acceptance
improvements, or any corrections. The Version of Record is available online at: http://dx.doi.org/10.1007/s10773-023-05369-x. Use of this Accepted Version is subject to the publisher’s Accepted
Manuscript terms of use https://www.springernature.com/gp/open-research/policies/acceptedmanuscript- terms}}
}
}}
}

\begin{document}

\newcommand{\be}{\begin{equation}} \newcommand{\ee}{\end{equation}}
\newcommand{\bea}{\begin{eqnarray}}\newcommand{\eea}{\end{eqnarray}}

\title{Quantum walk search  on a  two-dimensional grid with extra edges}

\author{Pulak Ranjan Giri} \email{pu-giri@kddi-research.jp}

\affiliation{KDDI Research, Inc.,  Fujimino-shi, Saitama, Japan}

\begin{abstract} 
Quantum walk has been   successfully used to  search for targets on  graphs with vertices   identified as  the  elements of   a database. 
This spacial search  on  a  two-dimensional periodic  grid  takes    $\mathcal{O}\left(\sqrt{N\log N}\right)$  oracle consultations     to find  a  target vertex   from $N$ number of vertices    with   $\mathcal{O}(1)$ success   probability, while reaching optimal speed of $\mathcal{O}(\sqrt{N})$ on   $d \geq 3$ dimensional square  lattice.  Our numerical analysis based on lackadaisical quantum walks searches    $M$  vertices  on  a 2-dimensional grid  with optimal speed of  $\mathcal{O}(\sqrt{N/M})$,  provided  the grid is attached with  additional long range edges.   Based on  the numerical analysis performed with multiple sets of randomly  generated  targets  for a wide range of $N$ and $M$ we suggest  that the  optimal  time complexity  of  $\mathcal{O}(\sqrt{N/M})$  with constant  success probability  can be achieved for quantum search on a   two-dimensional periodic grid with long-range edges.

 \end{abstract}


\date{\today}

\maketitle\thispagestyle{specialfooter}

\maketitle 



\section{Introduction} \label{in}
We are now in the era of   so called noisy intermediate scale quantum(NISQ) computing \cite{pres}, where we can   execute  small sized, up to a few hundred qubits,   quantum circuits in real quantum computer available today.  These devices are of limited capability because of its limited number of qubits with decoherence effects,  noisy gates  and without  the scope of  full-fledged error correction codes.  Nevertheless,  many interesting problems in the  hybrid quantum classical setting have been successfully  implemented  in NISQ device. However, the need for quantum computing   was understood long time ago  in the  $1980$s  by   Richard Feynman and  Paul Benioff, who thought that  computations based on the principles  of quantum mechanics   would be more efficient than classical computations.    The inefficiency of the classical  computation to simulate  some quantum systems even  with a modest size   together with the  ever decreasing size of the  transistor, leading to unavoidable  quantum effects,   motivated  researchers  to consider  quantum computations  as an alternative  form of   computing in near future.

There has been significant advances  in the theory of quantum computing  \cite{nielsen} to demonstrate the power, efficiency and speed   of quantum computer in comparison to classical computer.  For example,  Peter Shor's  \cite{shor1,shor2}  prime factorization algorithm   can  factorize   a large  number  into two prime numbers  in polynomial-time.   
Grover's  \cite{grover1, grover2,radha1,korepin1,zhang}   celebrated  quantum search  algorithm   can search  a target element in an unsorted database  in a time quadratically faster than  the exhaustive classical search.  To be more specific, Grover  quantum search takes  $\mathcal{O}(\sqrt{N})$ oracle consultations  \cite{giri}  as compared to  classical  search time  of  $\mathcal{O}(N)$  to find an  target element from an unsorted database of $N$ elements.

Since  Grover search has  applications as a subroutine of  several problems, its  generalisation     to   search  on spatial regions, e.g., on 
graphs \cite{childs,amba1,amba2,meyer,amba4} is an important aspect  of research. Straightforward application of Grover search on graph  shows it is not efficient though.   For example,  a  two-dimensional lattice   of size $\sqrt{N} \times \sqrt{N}$  requires   $\mathcal{O}(\sqrt{N})$   Grover iterations to search for a target vertex  and each iteration needs   $\mathcal{O}(\sqrt{N})$  unit of time to perform the reflection operation, making the total time  same as the classical running time  of $\mathcal{O}(N)$ \cite{beni}.   Recursive  algorithm \cite{amba1}  followed by amplitude amplification  
\cite{brassard}  can however search a target vertex   on   a  two-dimensional grid   in  $\mathcal{O}(\sqrt{N}\log^2 N)$ time, reaching optimal speed of   $\mathcal{O}(\sqrt{N})$  on   square lattices  with dimensions  more than two.   Another approach is to   search on   a graph  by  quantum walk(QW)  \cite{portugal}.   In this context, note that  the probability distribution  of quantum walk   spreads  quadratically  faster  than the probability distribution of classical  random walk.   Quantum walk  can search a vertex   on   a two-dimensional square lattice  in  $\mathcal{O} (\sqrt{N} \log N )$  time, reaching optimal time complexity of  $\mathcal{O} (\sqrt{N} )$     on    $d \geq 3$-dimensional  lattice   \cite{amba2,childs1}.

There has been several attempts to improve the  time complexity of  quantum search on   two-dimensional  square lattice.  An  improvement of  $\mathcal{O} ( \sqrt{\log N })$  \cite{tulsi, amba3, wong1}  in time complexity is possible  using different techniques  in quantum walk searching. 
However,    lackadaisical quantum walk  \cite{wong1,wong2,wong3}   can  search for a target element  in   $\mathcal{O} ( \sqrt{N \log N })$ time without the need for any additional technique.  Generalization   to   multiple target   vertices    \cite{rivosh,saha, nahi}  have  also been  investigated for several arrangements of the targets.  Lackadaisical quantum walk  can search one of the  $M$ vertices  in   $\mathcal{O} (\sqrt{N/M \log N/M })$ time \cite{giri2} with  constant  success probability.  It has been seen that  Hanoi network, which has extra long range edges,    have advantages on the  searching  capabilities \cite{boe,boe1,giri1}.    In this article  we  numerically  investigate the effects of extra long range edges   on the searching  efficiency  on  a   two-dimensional grid.  We consider   Hanoi network of degree four type extra edges on each line of both directions of the grid. Lackadaisical quantum walk search   can achieve the optimal speed of  $\mathcal{O} (\sqrt{N/M})$ in our  experiment.  The  scaling of  optimal time complexity  is supported by the numerical evaluations  performed on  multiple sets of randomly generated targets for a wide range of $N$ and $M$. The  comparison between the results of  two dimensional  grid with and without long range edges is  also performed. 
 
\begin{figure}[h!]
  \centering
     \includegraphics[width=0.80\textwidth]{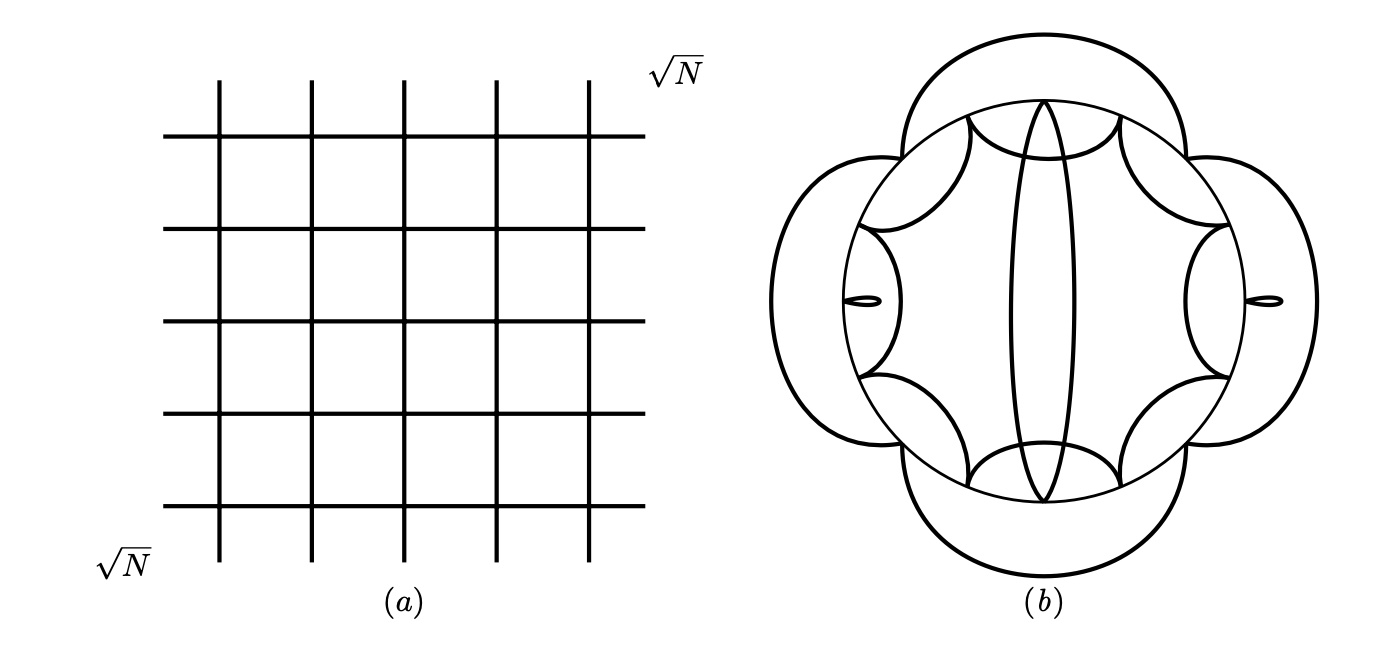}
          
       \caption{ (a)  Two dimensional grid of  size   $\sqrt{N} \times \sqrt{N}$   with periodic boundary conditions.   (b)  Each row and   column of  the grid in left panel  is additionally added with long-range edges of   Hanoi network of degree four  as  depicted in right panel for a $16\times 16$ square lattice.    A  self-loop  is also   added  at  each vertex  of the $2$-dimensional grid(left panel) for lackadaisical quantum walks.}
\end{figure}

We   arrange  this article  in the following fashion:  A brief discussion  on  quantum walk  search  is provided in section    \ref{qw}.   In section \ref{2D}   we discuss  lackadaisical quantum walk search on two-dimensional   periodic  grid  with extra long range edges  and  finally in section \ref{con} we conclude.

\section{ Quantum walk   search on  a graph} \label{qw}
In discrete time quantum walk, evolution from one vertex of  a graph to another nearest  neighbour vertex is made based on the state of the quantum coin.  To better express  it, let us consider a 
Cartesian graph $G(V, E)$  with  $V$ vertices and $E$ edges.   The vertices form the basis states of  the Hilbert space of vertices   $\mathcal{H}_V$  with dimensions  $N$, which is the number  of vertices on  the graph.  Similarly,  all the edges at each vertex form the  basis for the  coin space  $\mathcal{H}_C$ with dimensions $2d$ for a $d$-dimensional square lattice.
The Hilbert space  describing  the graph, $\mathcal{H}_G =  \mathcal{H}_C \times \mathcal{H}_V$,    is the tensor  product  space  of the  two spaces   $\mathcal{H}_V$   and  $\mathcal{H}_C$  respectively.   

\begin{figure}[h!]
  \centering
     \includegraphics[width=0.80\textwidth]{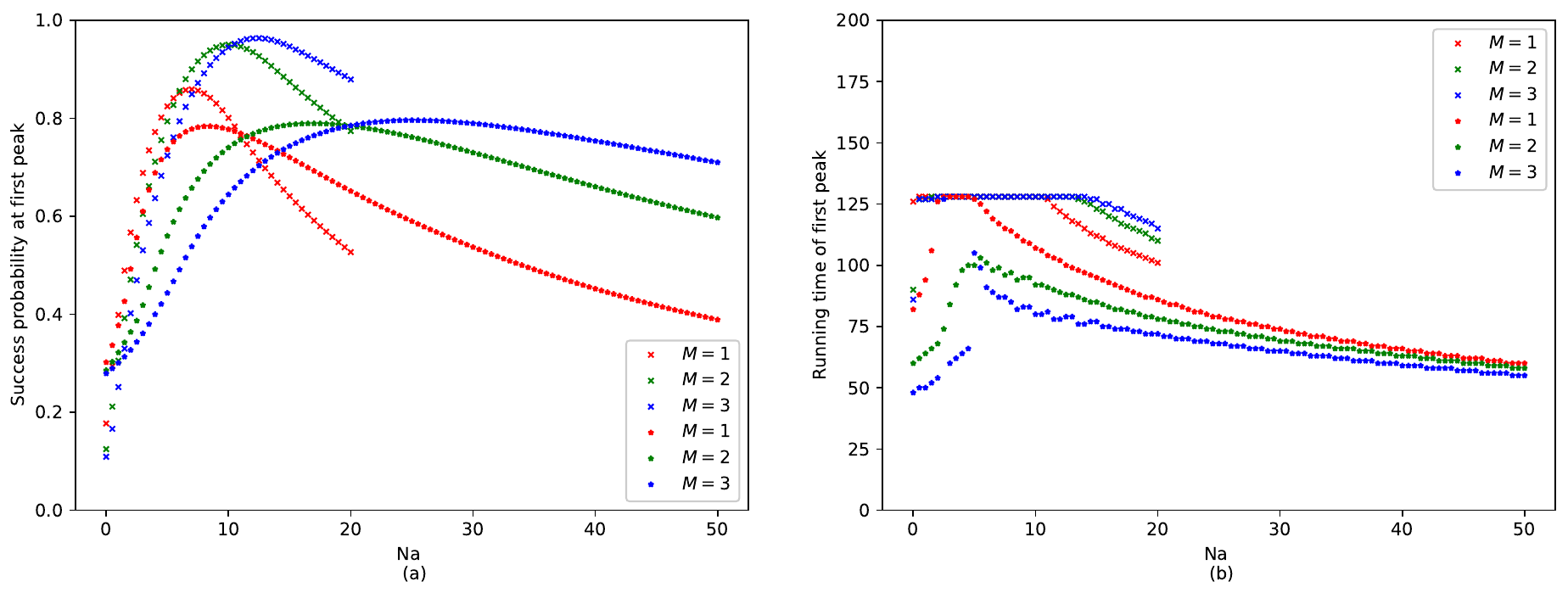}
          
       \caption{(Color online) (a) Variation of the first peak of the success probability  for  $M =1$(red), $2$(green) and $3$(blue) targets and (b) corresponding running time  as a function of the self-loop weight
       $Na$  for   a  $64\times 64$ square lattice.  $\times$-curves correspond to square lattice without long-range edges and $\star$-curves correspond to lattice with long-range edges respectively.
       }

\end{figure}

\begin{figure}[h!]
  \centering
     \includegraphics[width=0.80\textwidth]{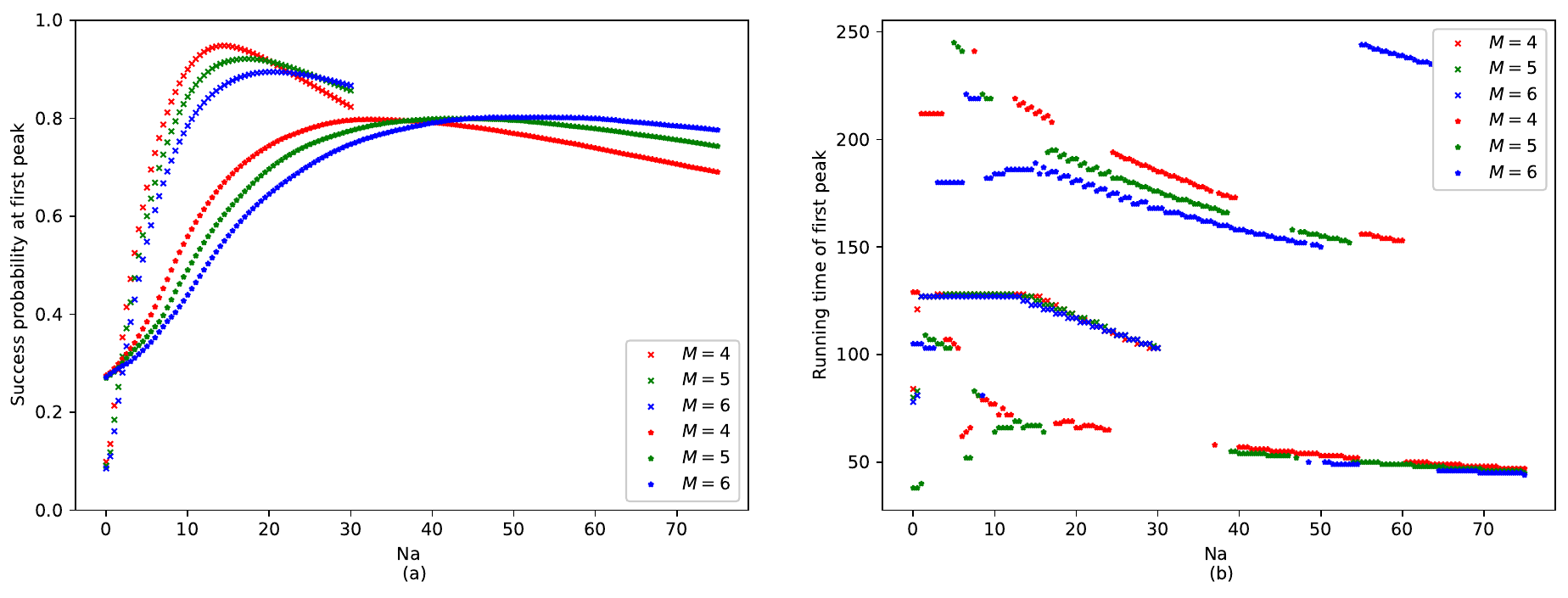}
          
       \caption{(Color online) (a) Variation of the first peak of the success probability  for  $M =4$(red), $5$(green) and $6$(blue) targets and (b) corresponding running time  as a function of the self-loop weight
       $Na$  for   a  $64\times 64$ square lattice.  $\times$-curves correspond to square lattice without long-range edges and $\star$-curves correspond to lattice  with long-range edges respectively.
       }

\end{figure}

In quantum walk search, we usually   start the evolution  with an initial  state which is equal superposition of the basis states.   The reason for the equal  superposition is that,  each basis state, which could be a possible  target state,  has the maximal overlap with the initial state.   The initial states  of the two Hilbert spaces are the following 
\begin{eqnarray}
 |\psi_c \rangle  &=& \frac{1}{\sqrt{2d}} \sum^{d}_{i =1} \left( |x_c^{i+}\rangle + |x_c^{i-}\rangle \right)\,,  \\
 |\psi_v \rangle  &=&   \frac{1}{\sqrt{N}} \sum^{\sqrt[d]{N}-1}_{x_v^i=0} |x_v^1,  x_v^2, \cdots,  x_v^d  \rangle \,,
\label{in}
\end{eqnarray}
where   $|\psi_c \rangle$ and  $|\psi_v \rangle$ are  the initial  states of the coin space and  vertex space respectively.   For the quantum walk search we have to evolve the initial state  of the graph,   $|\psi_{in}\rangle = |\psi_{in}\rangle \otimes  |\psi_{in}\rangle$,  by  a suitably chosen unitary operator $\mathcal{U}$   multiple times to achieve a constant probability for the target.   The operator    $\mathcal{U}$ is composed of  the evolution operator  for the coin and  shift operation  respectively.    However, for the quantum walk search standard shift operator fails to increase the success probability of the target state.  Fortunately,  there  exists other shift operator, known as the flip-flop shift operator,  $S$,  which can do the job of quantum walk search efficiently.    We also need to somehow mark the target state in the process of evolution, similar to what is done in Grover search.   It can be achieved by using  modified  coin operator,  $\mathcal{C}$,     as follows
\begin{eqnarray}
\mathcal{C}=  C \otimes \left( \mathbb{I} -   2   \sum_{i=1}^M |t_i \rangle \langle t_i |  \right)\,, 
\label{qc}
\end{eqnarray}
where    $C$ is the coin operator and     $|t_i \rangle$s  are  $M$   target states.   Note that, when the modified coin acts on the state involving the  target state  its phase becomes    negative, however when it acts on any other state, its phase  remains same.   We   choose  $C$ as the  Grover coin   for  our analysis.   The evolution operator for the quantum walk search   is  thus the following:
\begin{eqnarray}
 \mathcal{U} = S \mathcal{C}\,,
\label{uqw}
\end{eqnarray}
where the flip-flop shift operator   $S$  acts on both the vertex space and the coin space respectively.

The action of this shift operator is such that the quantum walker  moves from a given vertex to other nearest neighbour vertex  depending  on  the state of the  quantum coin  and then flips   the state of the coin.   In mathematical terms it is given by 
\begin{eqnarray} \nonumber
S =  \sum^{ \sqrt[d]{N}-1}_{x_v^i=0} \sum^{d}_{i=1}  |x_c^{i-} \rangle \langle x_c^{i+} | \otimes |x_v^1,  x_v^2, \cdots,  x_v^i+1, \cdots,  x_v^d  \rangle \langle x_v^1,  x_v^2, \cdots,  x_v^d  | +\\ 
  |x_c^{i+} \rangle \langle x_c^{i-} | \otimes |x_v^1,  x_v^2, \cdots,  x_v^{i}-1, \cdots,  x_v^d  \rangle \langle x_v^1,  x_v^2, \cdots,  x_v^d  |\,.
\label{fshift}
\end{eqnarray}
The final state    $|\psi_{f}\rangle =    \mathcal{U} ^{t} |\psi_{in }\rangle$  is obtained after repeated  application,  $t$ times,  of  the evolution operator.   If the overlap  of the final   state with the  target  state  is high, $ |\langle  t_v |\psi_{f }\rangle| > 1/2$, and constant,  we have found the target state with time complexity   $t$.  However, if the overlap  is 
 $ |\langle t_v |\psi_{f }\rangle|    << 1 $,    we  need to  apply  amplitude  amplification   technique     $ 1/ |\langle t_v |\psi_{f }\rangle|$   times   on the final state  $|\psi_{f}\rangle$ 
 to achieve   $\mathcal{O}(1)$   success probability.    Time complexity  for  the search  with  quantum walk  followed by the amplitude amplification  thus becomes    $t/ |\langle t_v |\psi_{f }\rangle|$.
 
\begin{figure}[h!]
  \centering
     \includegraphics[width=0.80\textwidth]{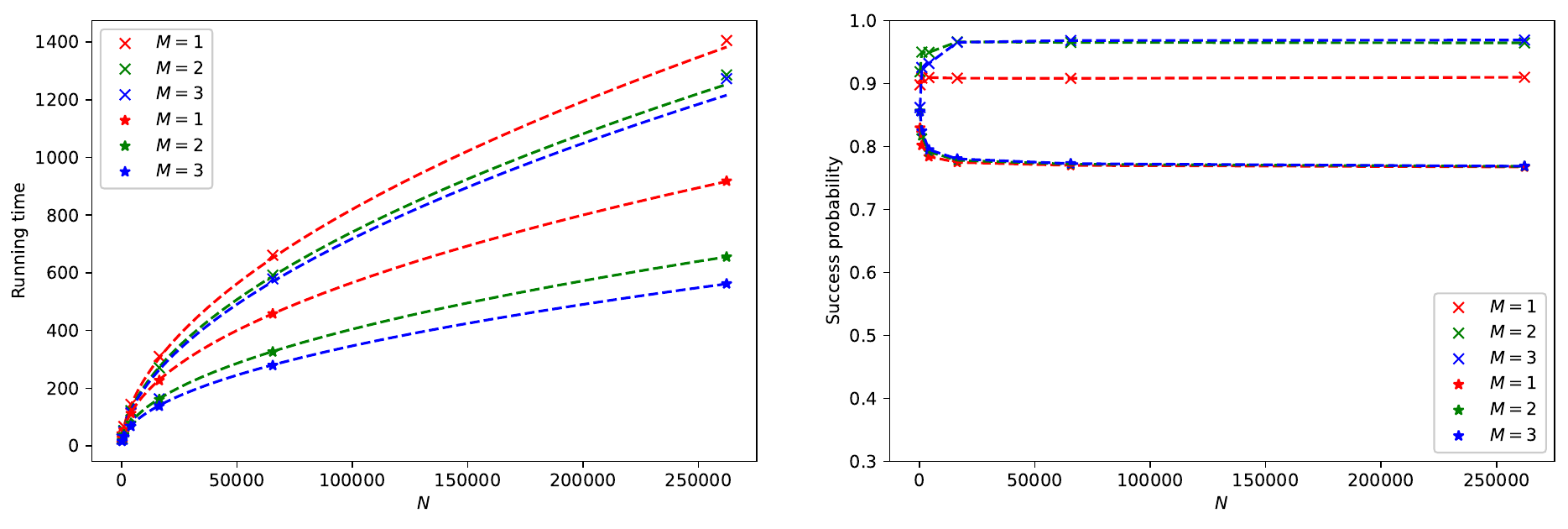}
          
       \caption{(Color online) (a) Running time of lackadaisical quantum walk search for $M =1$(red), $2$(green)  and $3$(blue) randomly chosen   targets on a two dimensional grid with and without  extra long-range edges  and (b)  the  corresponding success probability    as a function of  number of elements  $N$.  $\times$-curves correspond to square lattice without long-range edges and $\star$-curves correspond to lattice with long-range edges respectively.}

\end{figure}

\begin{figure}[h!]
  \centering
     \includegraphics[width=0.80\textwidth]{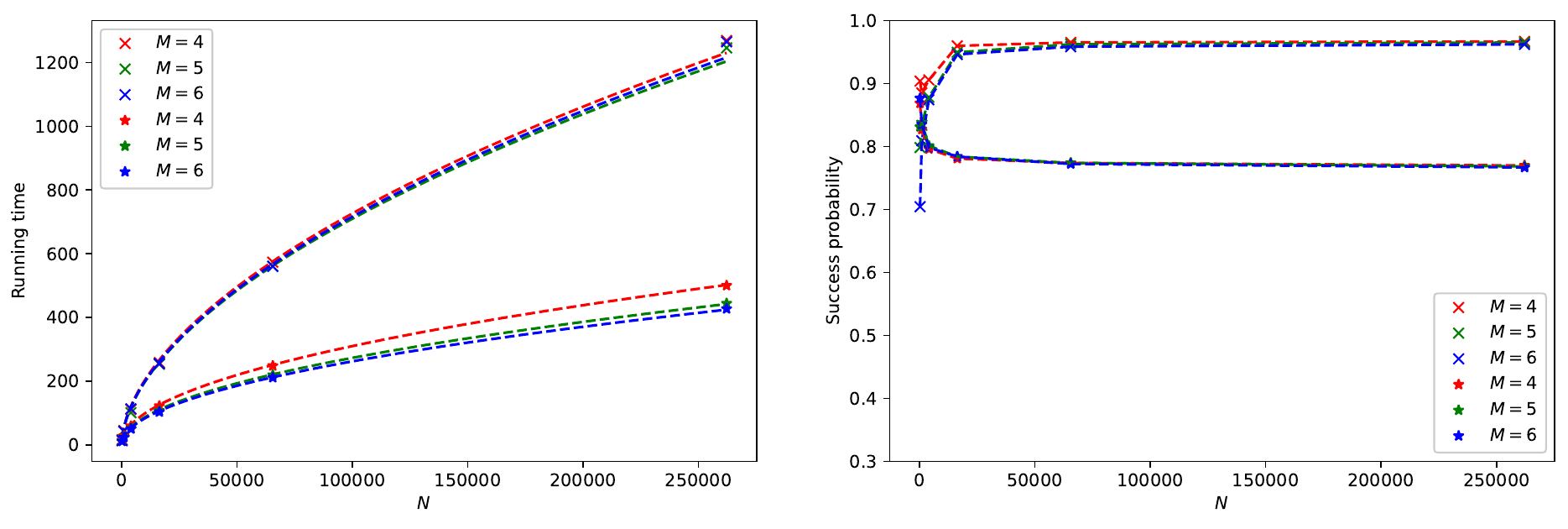}
          
       \caption{(Color online) (a) Running time of lackadaisical quantum walk search for $M =4$(red), $5$(green)  and $6$(blue)  randomly chosen  targets on a two dimensional grid with and without  extra long-range edges  and (b)  the  corresponding success probability    as a function of  number of elements  $N$.  $\times$-curves correspond to square lattice without long-range edges and $\star$-curves correspond to lattice with long-range edges respectively.}

\end{figure}

\section{Quantum search with   long range edges} \label{2D}
In this section we discuss  how we can search for target vertices  on a  $2$-dimensional periodic  grid  with additional  long range edges with lackadaisical quantum walk.  In FIG. 1(a)  a   periodic $2$-dimensional   grid     of size   $\sqrt{N} \times \sqrt{N}$ is displayed, where    $N$    elements of  a database are represented  as  the   $N$   vertices     $(x, y)$   of the graph.   Each  vertical and horizontal line of the graph   is associated with long range edges as can be seen from an example graph given  in FIG. 1(b)  which can be  associated with a $16\times 16$  square lattice.  The  long range edges are formed in accordance with the conditions  of the  Hanoi network of degree four(HN4) \cite{boe,boe1,giri1}, which plays a crucial role in quantum search.   Each vertex is connected to  nine  edges, four of them come from the four edges of the $2$-dimensional grid,  other four come from the long range edges of the  two HN4 networks  and one comes from the self-loop of lackadaisical quantum walk.   Therefore,   initial state of the quantum coin  is given by:
\begin{eqnarray}
|\psi_c \rangle   =  \frac{1}{\sqrt{8+ a}} \left( |x_c^{0+}\rangle + |x_c^{0-}\rangle+ 
 |x_c^{1+}\rangle + |x_c^{1-}\rangle  + |x_c^{2+}\rangle + |x_c^{2-}\rangle+ 
 |x_c^{3+}\rangle + |x_c^{3-}\rangle   + \sqrt{a}| x_c^0\rangle \right) \,.
\label{2Dcstate}
\end{eqnarray}
In terms of the coordinates of  the HN4 network each vertex of the $2$-dimensional grid   $ 1 \leq x_v^1, x_v^2 \leq 2^n$  is located    at  position  $(i_1, i_2)$   in a specific hierarchy   $(j_1,j_2)$, which obey   the following  relations 
\begin{eqnarray} \nonumber
x_v^1 &=& 2^{i_1} \left(  2j_1 +1  \right)\,,\\
x_v^2 &=& 2^{i_2} \left(  2j_2 +1  \right)\,,
\label{hier}
\end{eqnarray}  
where  $0 \leq i_1, i_2 \leq n$,   $0 \leq j_1 \leq {j_1}_{max} = \lfloor2^{n-i_1-1} -1/2 \rfloor$ and  $0 \leq j_2 \leq {j_2}_{max} = \lfloor2^{n-i_2-1} -1/2 \rfloor$  uniquely identifies the vertex in the $2$-dimensional grid.   The basis  states   for the grid 
\begin{eqnarray}
|x^1_v, x^2_v\rangle  =   |i_1, j_1; i_2, j_2 \rangle\,,
\label{hier}
\end{eqnarray}  
can be   expressed   in terms of  the Cartesian coordinates   $x^1_v, x^2_v$ or   alternatively   in terms of the HN4 coordinates   $i_1, j_1; i_2, j_2$  respectively.     
The initial state of  the  vertex space of the graph  is given by 
\begin{eqnarray}
|\psi_{v}\rangle =   \frac{1}{\sqrt{N}} 
\sum^{\sqrt{N}}_{{x^1_v, x^2_v} = 1} |x_v^1,  x_v^2\rangle \,.
\label{2Din}
\end{eqnarray}
In our quantum search problem   we choose the Grover diffusion operator  
\begin{eqnarray}
C = 2 |\psi_c \rangle \langle \psi_c | - \mathbb{I}_9\,, 
\label{2Dcgrov}
\end{eqnarray}
for  the rotation of the  quantum coin state.    The shift operator,  $S = S_1 + S_2 + S_3$,  is composed of  three  parts,  which are associated   with   three different types of the edges.  The part  associated with  four standard edges of the $2$-dimensional grid is  given by
\begin{eqnarray} \nonumber
S_1 =  \sum^{ \sqrt{N}}_{x^1_v, x^2_v=1}  && \left[ |x_c^{0-} \rangle \langle  x_c^{0+} | \otimes | x^1_v +1, x^2_v  \rangle \langle  x^1_v, x^2_v  | 
+   |x_c^{0+} \rangle \langle  x_c^{0-} | \otimes | x^1_v -1, x^2_v  \rangle \langle  x^1_v, x^2_v  |     \right. \\
 &+& \left.  |x_c^{1-} \rangle \langle  x_c^{1+} | \otimes | x^1_v, x^2_v  +1 \rangle \langle  x^1_v, x^2_v  |  
+   |x_c^{1+} \rangle \langle  x_c^{1-} | \otimes | x^1_v, x^2_v -1 \rangle \langle  x^1_v, x^2_v  |     \right]\,,  
\label{fshift1}
\end{eqnarray}
where the coin  basis states  $|x_c^{0+} \rangle, |x_c^{0-} \rangle, |x_c^{1+} \rangle, |x_c^{1-} \rangle $ are  associated with  right, left, up and down edges respectively.
The shift operator associated with four long range edges  at each vertex is given by 
\begin{eqnarray} \nonumber
S_2  &=&\sum^{ \sqrt{N}}_{x^2_v=1} \sum^{ n-2}_{i_1=0}   \sum^{{j_1}_{max}}_{j_1=0}  \left(  |x^{2-}_c \rangle \langle x^{2+}_c | \otimes | i_1, j_1 + 1; x^2_v  \rangle \langle  i_1, j_1; x^2_v |  + |x^{2+}_c \rangle \langle x^{2-}_c | \otimes | i_1, j_1 - 1; x^2_v  \rangle \langle  i_1, j_1; x^2_v | \right) \\ \nonumber 
&+& \sum^{ \sqrt{N}}_{x^1_v=1} \sum^{ n-2}_{i_2=0}   \sum^{{j_2}_{max}}_{j_2=0} \left(  |x^{3-}_c \rangle \langle x^{3+}_c | \otimes | x^1_v; i_2, j_2 + 1  \rangle \langle x^1_v; i_2, j_2 |  
+ |x^{3+}_c \rangle \langle x^{3-}_c | \otimes | x^1_v; i_2, j_2 - 1  \rangle \langle  x^1_v; i_2, j_2 | \right) \\ \nonumber 
&+& \sum^{ \sqrt{N}}_{x^2_v=1} \left( | x^{2-}_c \rangle \langle x^{2+}_c | +   | x^{2+}_c \rangle \langle  x^{2-}_c |  \right)\otimes  \left( | n-1 , 0; x_v^2  \rangle \langle  n-1, 0;x^2_v |  +   | n , 0; x^2_v  \rangle \langle  n, 0; x^2_v | \right)  \\
&+& \sum^{ \sqrt{N}}_{x^1_v=1} \left( | x^{3-}_c \rangle \langle x^{3+}_c | +   | x^{3+}_c \rangle \langle  x^{3-}_c |  \right)\otimes  \left( | x^1_v; n-1 , 0  \rangle \langle  x^1_v;n-1, 0 |  +   | x^1_v; n , 0  \rangle \langle  x^1_v; n, 0 | \right)\,,
\label{fshift2}
\end{eqnarray}
where the coin  basis states  $|x_c^{2+} \rangle, |x_c^{2-} \rangle, |x_c^{3+} \rangle, |x_c^{3-} \rangle $ are  associated with  right, left, up and down long range edges respectively.
And finally the shift operator associated with  the  self-loop of   lackadaisical quantum walk is given by 
\begin{eqnarray} 
S_3  =  \sum^{ \sqrt{N}}_{x^1_v,x^2_v=1}    |x^{0}_c \rangle \langle x^{0}_c | \otimes | x^1_v; x^2_v  \rangle \langle  x^1_v; x^2_v | \,.
\label{fshift3}
\end{eqnarray}
There are nine possible directions a quantum walker can take at each vertex.   Assuming the walker being at $|x^1_v; x^2_v\rangle$ it can  go to  one  of  $|x^1_v \pm 1; x^2_v\rangle$, $|x^1_v; x^2_v \pm 1 \rangle$ if the coin states are among  the four edges of the grid or it can go to one of  $|i,j  \pm 1; x^2_v\rangle$, $|x^1_v; i,j \pm 1 \rangle$ if the coin states are among the four long range edges.  When the coin state is the self-loop of the lackadaisical quantum walk then the walker stays  at the same vertex.  There are  undirected self-loops   for  $i=n-1, n$, i.e. at  $|n-1,0; x^2_v\rangle$,  $|n,0; x^2_v\rangle$,
$|x^1_v; n-1,0 \rangle$ and  $|x^1_v; n,0 \rangle$, which are exceptional points.

\begin{table}
    \begin{tabular}{ |  p{1cm }  | p{2cm } |  p{2cm} | p{3.3cm} | p{3.0cm} | p{3.0cm} |}
    \hline
    \textbf{M targets}   & \textbf{Na(without  long-range edges)}  & \textbf{Na(with long-range edges)} &  \textbf{Target states} & \textbf{Running time(without  long-range edges)} & \textbf{Running time(with long-range edges)} \\ \hline\hline
    
    1 & $7.0$  & $8.5$.  &  $|1;6\rangle$   &   $1.16\sqrt{N\log N}$  &   $1.79\sqrt{N}$\\ \hline
    
    2  &   $10.0$  &  $17.0$  &   $|11;1\rangle, |9;12\rangle$  &  $1.53\sqrt{N/2\log N/2}$  &  $1.81\sqrt{N/2}$ \\ \hline

    3 &  $12.5$  &   $25.0$  &  $|4;10\rangle, |14;8\rangle, |0;12\rangle$  & $1.85\sqrt{N/3\log N/3}$  & $1.90\sqrt{N/3}$  \\  \hline

    4  &  $14.5$ &  $32.0$   &  $|11;0\rangle, |2;5\rangle, |3;12\rangle$, $|8;9\rangle$ &  $2.19\sqrt{N/4\log N/4}$  &  $1.96\sqrt{N/4}$  \\  \hline

    5  & $17.5$  &   $44.5$   &  $|12;13\rangle, |1;4\rangle, |9;8\rangle$, $|2;11\rangle, |14;6\rangle$ &  $2.42\sqrt{N/5\log N/5}$  &  $1.93\sqrt{N/5}$ \\  \hline

    6  &  $20$  &  $53.0$  & $|2;9\rangle, |6;13\rangle, |6;7\rangle$, $|10;11\rangle, |4;5\rangle, |0;14\rangle$ & $2.70\sqrt{N/6\log N/6}$  & $2.03\sqrt{N/6}$ \\  \hline
    
 \end{tabular}
 \caption{Optimal self-loop weights for six quantum searches with randomly generated  targets spread over the  size  of  $16 \times 16$ excluding the exceptional points   and  corresponding running time for two-dimensional grid  of size up to $512 \times 512$ with and without long-range edges.} 
 \end{table} 
To search by  lackadaisical quantum walk  we need to choose a suitable value for the   self-loop  weight.  Usually self-loop weight $a$ depends  the number of elements in the database $N$, number of targets $M$, the type of graph under study and dimensions of the graph.  In our case the graph and its dimensions are fixed as we are working with  only   two-dimensional grid.   The  self-loop  weight  depends on the   number of elements $N$  as  $a \propto 1/N$, where  $Na$ depends on the number of targets  to search in the database.  So we  find the optimal value  for a fixed $N$ for a fixed number of  targets. 
In FIG. 2 the behaviour of the success probability of the first peak  and corresponding running time  with respect to the weight $Na$  for a  $64\times 64$ square grid is presented.  We have considered three different target configurations, namely   $M =1$(red),  $M =2$(green),  and  $M =3$(blue) targets  with  random target states.  The value of the self-loop weights which correspond to the peaks of the  six  curves  in FIG. 2(a) are  chosen as the optimal values  for the study of quantum walk search.  The summary of optimal self-loop weights and corresponding states are  given in Table 1. Similar analysis has also been performed for
$M =4$(red),  $M =5$(green),  and  $M =6$(blue) targets as can be seen from  FIG. 3.

\begin{figure}[h!]
  \centering
     \includegraphics[width=0.80\textwidth]{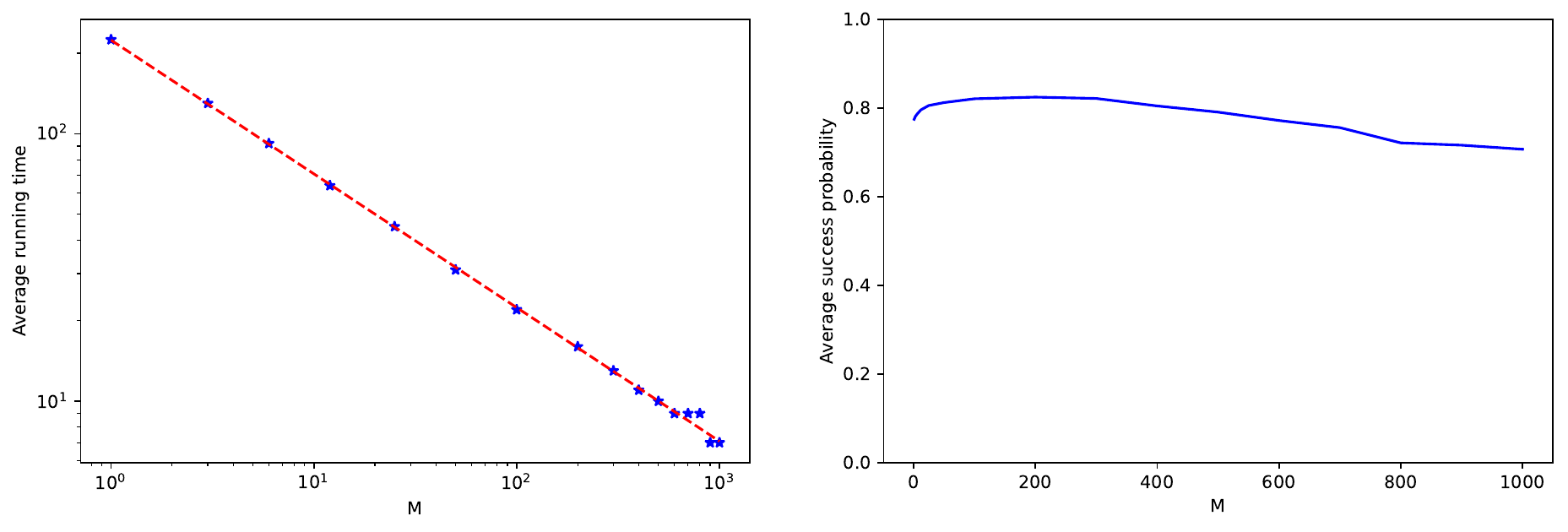}
          
       \caption{(Color online) (a) Average running time of lackadaisical quantum walk search on  a set of randomly generated   targets  ranging in the interval $M \in [1,1000]$ on a $128 \times 128$ two-dimensional grid with  extra long-range edges  and (b)  the  corresponding average  success probability    as a function of  number of targets  $M$. Average is calculated on a set of $50$  randomly chosen targets at fixed $M$.    $\star$-curve corresponds to experimental data on running time and dashed red-curve  corresponds to  $1.75\sqrt{N/M}$. The sets of targets are randomly chosen excluding the exceptional points, where long-range edges form  undirected self-loops.  Self-loop weight for the lackadaisical quantum walk $Na= 8.5M$.}

\end{figure}

\begin{figure}[h!]
  \centering
     \includegraphics[width=0.40\textwidth]{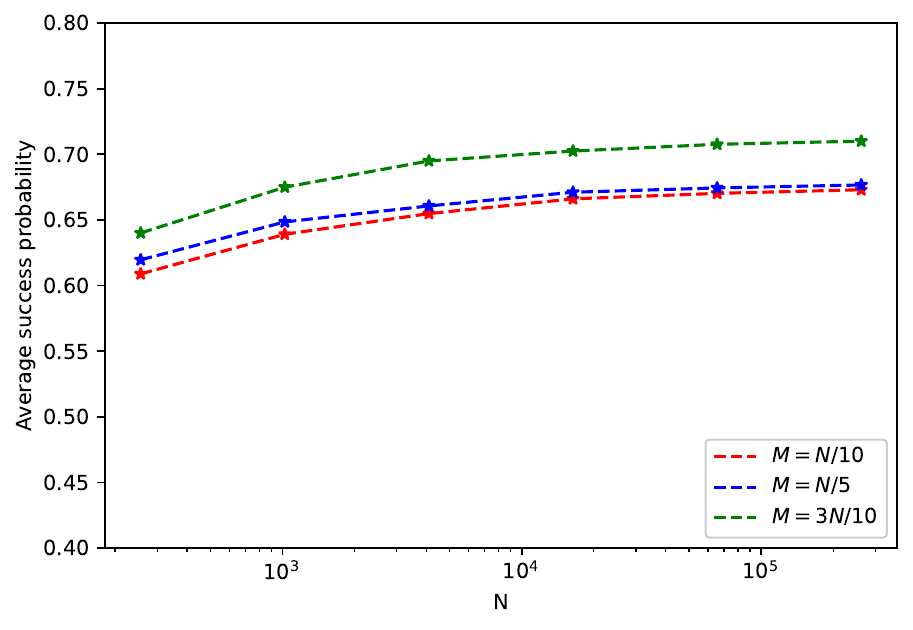}
          
       \caption{(Color online) (a) Average success probability  over a set of  randomly  generated  $50$ targets as a function on the lattice size $N$.  Red, blue and green  curves  correspond to cases with 
       $10 \%$, $20 \%$ and $30 \%$  of the vertices  are marked   respectively.   Self-loop weight for the lackadaisical quantum walk $Na= 8.5M$.}

\end{figure}

Numerical analysis for the running time and success probability for lackadaisical quantum walk search is performed on a two-dimensional grid of size  up to   $512 \times 512$.  Both grid with and without long-range edges are considered for the analysis.  In FIG. 4   $M=1, 2$ and $3$  randomly generated  targets are used for the searching purpose.  $\times$-marked curves   correspond to the standard  quantum walk search without long-range edges, which have success probability  $\sqrt{\log N/M}$ times the success probability  for the case with long-range edges represented by  $\star$-marked  curves.   For  $M=4, 5$ and $6$ randomly generated   targets,  result  for the running time and success probability  is presented in FIG. 5.  Our analysis as summarised in Table 1,  suggests   that the time complexity  for the lackadaisical quantum walk search on a two dimensional grid  with extra long range edges  is    
\begin{eqnarray}
t =  \mathcal{O} \left(\sqrt{N/M} \right),
\label{tcom}
\end{eqnarray}  
which is the optimal time complexity for searching for targets in an unsorted database.  The coefficient for the running times  in   fifth and sixth column of Table 1 have  been obtained by curve-fitting(dashed lines in running time) the  data obtained from  the numerical analysis.    Further analysis on wide ranges of $N$ and $M$ performed in   FIG.  6 and. FIG. 7  supports in favour of the validity  of  the time complexity and success probability   presented in Table 1  for two-dimensional periodic grid with long range edges.  In FIG. 6 
average running time and success probability is evaluated   over a  set of  $50$ randomly generated targets(excluding exceptional points)  with  $M$  ranging in the interval $M \in [1,1000]$ on a $128 \times 128$ two-dimensional grid with  extra long-range edges.   

Running time scales as   $1.75\sqrt{N/M}$.   In FIG. 7   success probability is calculated  for lattice size up to $512 \times 512$ for randomly generated targets for three settings where $10 \%$, $20 \%$ and $30 \%$ vertices of the lattice are target vertices respectively with running time  $\mathcal{O} \left(\sqrt{N/M} \right)$.  
As we can see  the success probability is well above $50 \%$ in all the three cases.

The vertices  at  $|n-1,0; x^2_v\rangle$,  $|n,0; x^2_v\rangle$,  $|x^1_v; n-1,0 \rangle$ and  $|x^1_v; n,0 \rangle$ are the exceptional points on the grid which can not be searched  by quantum walk because at these points the long-range edges form undirected self-loops, which prohibit the probability  flux to flow-in  from other neighbouring  vertices through long-range edges. 

\section{Conclusions} \label{con}
Grover quantum search is one of the most important quantum algorithms, which can find $M$ targets  from an unsorted database of $N$ elements in  $\mathcal{O}(\sqrt{N/M})$ time steps.  This is the optimal time complexity \cite{zalka} for searching as no other algorithm for searching  can be faster than the Grover algorithm.  The spacial search  on  a  two-dimensional periodic  grid, on the other hand,   takes    $\mathcal{O}\left(\sqrt{N/M\log N/M}\right)$  time steps     to find  $M$  vertices  from the graph with size  $\sqrt{N} \times \sqrt{N}$.   Optimal speed of $\mathcal{O}(\sqrt{N/M})$ is achieved on $d \geq 3$ dimensional square  lattice.  Even increasing the degree of the vertices of the two-dimensional  lattice  does not improve the running time.  For example, honeycomb, rectangular and triangular grid with degree  $3, 4$ and $6$ respectively  have the same running time \cite{nahi}. 

However, we  in this article showed that by adding extra long-range edges  of the type of  Hanoi network of degree four  with each horizontal and vertical lines of the grid and using lackadaisical quantum walk we  can search some randomly generated   targets, presented in Table 1,  in optimal  time $\mathcal{O} (\sqrt{N/M})$. Numerical  simulations  on  database size of up to $N = 512\times 512=  262144$  elements   have  been performed with up to $M=6$   targets. Best fitting curves with the obtained data  from numerical analysis  show that the time complexity is optimal  for randomly generated  targets,  as can be seen from  the  right most column in Table 1.  Based on the numerical analysis we suggest that the time complexity for searching  on a two-dimensional grid with extra long-range edges is  $\mathcal{O} \left(\sqrt{N/M} \right)$. 
The reason for this faster running time is that the long-range edges   allow for additional  and faster flow of probability flux.

There are some vertices, which are not connected by long-range edges with neighbouring vertices, because on those  vertices long-range edges form self-loop. These  exceptional vertices at   $|n-1,0; x^2_v\rangle$,  $|n,0; x^2_v\rangle$,  $|x^1_v; n-1,0 \rangle$ and  $|x^1_v; n,0 \rangle$ can not be searched by our quantum walk search algorithm. It would be interesting to investigate the effects of long-range edges on quantum search on other graphs also.   It would be  nice  if the numerical result presented in this article  can be  obtained by   analytical method. 

\vspace{1cm}

Data availability Statement:  The datasets generated during and/or analysed during the current study are available from the corresponding author on reasonable request.
\vspace{0.5cm}

Conflict of interest: The authors have no competing interests to declare that are relevant to the content of this article.


\end{document}